**Detachment limited interlayer transport processes during SrTiO$_3$ pulsed laser epitaxy.**

Jeffrey G. Ulbrandt, Xiaozhi Zhang*, and Randall L. Headrick[†]

Department of Physics and Materials Science Program, University of Vermont, Burlington, Vermont 05405, USA

* Present address: X-ray Science Division, Advanced Photon Source, Argonne National Laboratory, Lemont, IL, 60439, USA.

[†]Corresponding author: rheadrick@uvm.edu

## ABSTRACT

Pulsed laser epitaxial growth is characterized by high instantaneous deposition rates that leads to the nucleation of transient islands, both on surface terraces, and on top of stable islands formed during previous deposition pulses. We report results from combined in-situ X-ray reflectivity and kinetic Monte Carlo (kMC) simulations. Specular reflectivity monitors interlayer transport, while diffuse scattering reveals the evolution of in-plane length scales, both during the recovery time between individual laser pulses and over multiple deposited layers. The initial stage after each laser pulse is faster than the temporal resolution of the experiment, while subsequent recovery occurs over seconds. The results suggest that transient islands on top of stable two-dimensional islands form immediately after the deposition pulse, and then ripen via detachment and diffusion, leading to the slower component. kMC simulations show that the detachment energy barrier plays a dominant role in determining the recovery time constant.

## I. INTRODUCTION

Pulsed laser deposition (PLD) is a versatile technique for the epitaxial growth of complex oxides, such as $SrTiO_3$. Its inherently high instantaneous deposition rates enable the formation of unique thin-film structures and the stabilization of complex oxide phases that are often inaccessible by other methods, establishing PLD as a powerful method for advancing materials discovery and exploring emergent physical phenomena.[1] Notable examples include the formation of a high-mobility electron gas at the at the $LaAlO_3/SrTiO_3$ heterointerface[2] and superconductivity in nominally insulating oxide heterostructures.[3] The PLD process encompasses both extremely fast (possibly sub-microsecond-scale) non-thermal relaxation events and slower (millisecond to second-scale) thermally driven relaxation processes that govern interlayer transport and intralayer rearrangements.

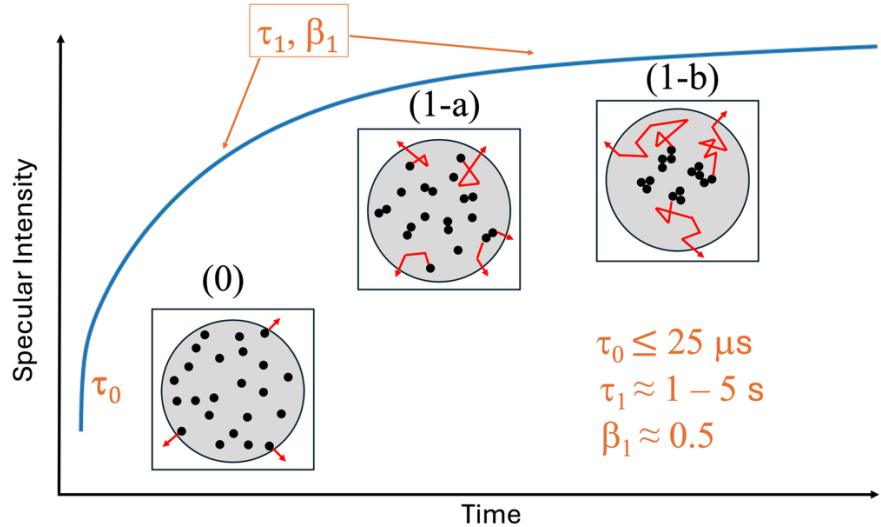

**Figure 1.** Schematic of the X-ray specular reflectivity recovery profile following a laser pulse. There are two regimes, one fast regime characterized by time constant $\tau_0$, and the other slow regime characterized by a stretched exponential function with time constant $\tau_1$ and stretching exponent $\beta_1$. The insets show trajectories of deposited monomers on underlying two-dimensional crystalline islands, from left to right: (*0*) nonthermal interlayer transport of monomers, (*1-a*) detachment from small transient islands and diffusion of monomers or several-atom clusters, and (*1-b*) detachment from larger transient islands.

Time-resolved X-ray diffraction studies by Tischler et al. revealed that fast, nonequilibrium interlayer transport can dominate during plume arrival, preceding slower thermal processes.[4] Fleet et al.,[5] further demonstrated that step-edge density influences surface smoothing by facilitating incorporation of energetic species. Simulations by Vasco and Sacedón suggest that hot clusters may exhibit ballistic mobility due to non-thermal kinetic energies.[6] This regime is labelled (0) in Figure 1.

Thermal relaxation dynamics in PLD share similarities with those observed in continuous deposition techniques like molecular beam epitaxy. Kaganer et al. demonstrated that post-growth annealing of GaAs involves two distinct recovery stages: thermally driven interlayer transport that fills ad-vacancies, and a slower in-plane coarsening of two-dimensional islands.[7] In $SrTiO_3$ PLD, analogous mechanisms are at work, although the coarsening dynamics can be significantly suppressed or vary with island size.[8]

In this work, we focus on the high nucleation density of unit-cell-height islands that form immediately upon plume arrival.[9] During the dwell period following each deposition pulse, these transient islands undergo a form of two-dimensional Ostwald ripening; smaller islands tend to shrink as their material detaches and diffuses, while larger islands grow via step-edge attachment.[10] However, we note that classic Ostwald ripening occurs in a closed system where mass is conserved within a fixed ensemble of particles[11,12] In contrast, the PLD growth surface is an open system, with continuous downward interlayer transport that dynamically alters the island size distribution. Although non-thermal processes may be important at very early times, we focus on thermally driven processes that dominate on millisecond and later time scales.

Our experimental results reveal two overlapping components of *thermally driven* interlayer relaxation labelled (*1-a*) and (*1-b*) in Figure 1. The first, occurring on a ~0.1 s timescale, arises from particles (either atoms or few-atom clusters) that directly diffuse to step edges without becoming trapped in transient islands. The second, slower component—on the order of seconds—is associated with particles that initially attach to transient islands and later detach before reaching lower-terrace step edges. Notably, the combined process deviates from simple exponential relaxation kinetics.

Figure 1 presents a schematic of the specular anti-Bragg X-ray reflectivity recovery following a laser pulse. The main panel shows two regimes of relaxation: a fast nonthermal component characterized by time constant $\tau_0$, and a slower component described by a stretched exponential with a time constant $\tau_1$ and stretching exponent $\beta_1$. Insets illustrate three conceptual stages: (*0*) fast interlayer transport of monomers; (*1-a*) thermally driven interlayer transport of atoms or small clusters; and (*1-b*) detachment from transient islands, where higher detachment barriers can delay relaxation. These processes define the distinct temporal regimes. Stage *0* is attributed to non-thermal motion immediately following the deposition pulse.[4,5,13] Monomers landing near island edges can hop down via non-thermal pathways, while those near the interior become thermalized. Stage *1* encompasses two overlapping sub-processes: *1-a*, in which newly deposited particles or small clusters diffuse directly to step edges; and *1-b*, in which particles that become trapped on transient islands are subsequently re-emitted on longer timescales. The relaxation profile may be described by a single stretched exponential when detachment barriers are modest but diverges into two distinct timescales when the barrier is large. Stretched exponential kinetics are common in systems with multiple competing pathways and a distribution of relaxation processes.[14] For coverages exceeding 0.5 monolayers, islands may evolve into holes, reflecting the interplay between island coalescence and thermally activated coarsening.

The solid-on-solid model[15] provides a useful framework for describing these complex processes by incorporating mechanisms such as surface diffusion, step-edge diffusion, corner diffusion, and step-edge detachment. We implement the SOS model in a computationally efficient kinetic Monte Carlo (kMC) algorithm that explicitly includes atomistic processes such as surface diffusion, attachment, and detachment. This allows us to simulate the formation and dissolution of transient islands and the resulting interlayer transport. To support the kMC results, we also consider continuum models that captures some features of the relaxation behavior but neglects the transient attachment mechanism.

Several rate-equation models have been proposed to describe the evolution of surface roughness and island size distributions during pulsed epitaxial growth. Among them, the model introduced by Vasco, Polop, and Sacedón (VPS) incorporates Ostwald ripening by distinguishing between subcritical two-dimensional clusters—which permit perfect interlayer transport—and supercritical islands, where interlayer transport is blocked by strong step-edge barriers.[16] Although VPS captures coarsening behavior, it fails to account for the recovery of specular X-ray reflectivity in PLD due to its omission of interlayer mass transfer from material deposited atop larger islands. As a result, thermally driven disintegration of transient islands—a key mechanism in our work—is entirely absent. Instead, clusters atop stable islands evolve through conventional Ostwald ripening, leading to coarsening *without* dissolution, thermally driven interlayer transport, or recovery of the specular intensity between deposition pulses.

Other rate-based models, such as that proposed by Xu et al., include deposition, intralayer diffusion, interlayer transport, and island dissolution driven by Gibbs–Thomson and Ostwald ripening effects.[17] In this formulation, an empirical parameter $\eta$ governs the rate at which upper-layer islands disintegrate when the underlying terrace is incompletely filled. While this model captures the transient nature of clusters atop stable islands, it assumes a constant dissolution probability, resulting in predictions of simple exponential relaxation that are inconsistent with experimental data for layer-by-layer growth during PLD.[5]

In contrast, our kMC simulations explicitly resolve thermally activated detachment from transient islands and naturally reproduce the stretched-exponential relaxation observed in specular X-ray reflectivity. We find that it is essential to include the formation and decay of transient clusters to accurately model the associated interlayer transport process. The simulations also predict diffuse scattering features that agree well with experiment, confirming the role of detachment-limited interlayer transport in shaping PLD growth dynamics.

While kMC methods offer detailed insights into surface growth dynamics, they are computationally intensive, particularly when simulating thermally activated processes over experimentally relevant time and length scales. As a result, previous studies have frequently focused on submonolayer growth regimes ,[9,18-20] or have introduced physical approximations to accelerate simulation times.[21] In this work, we implement a full kinetic Monte Carlo approach based on the solid-on-solid (SOS) approximation using a conventional bond-counting scheme,[22] without further assumptions. This allows us to model not only specular reflectivity recovery but also the X-ray diffuse scattering—an essential capability largely absent in previous computational studies.

## II. METHODS

### A. Sample Preparation and PLD Conditions

SrTiO$_3$ (STO) substrates were obtained from CrysTec GmbH and prepared by etching in buffered NH$_4$F-HF to produce a TiO$_2$-terminated surface, followed by high-temperature annealing to form atomically smooth terraces. Pulsed laser deposition (PLD) was carried out using a 248 nm excimer laser under an oxygen partial pressure of 2 mTorr to ensure proper oxidation during growth. The deposition rate was approximately 18 pulses per unit cell, with a dwell time of 6 seconds between pulses.

### B. In-Situ X-Ray Scattering

Real-time X-ray scattering measurements were performed at the Integrated In-situ and Resonant Hard X-ray (ISR) beamline at the National Synchrotron Light Source II. Scattering was recorded both on and off the specular condition using 11.3 keV photons to obtain in-plane and out-of-plane structural information during film growth. Data were collected at the [0 0 ½] anti-Bragg position to maximize sensitivity to single-unit-cell-height features.

Scattering patterns were captured using an Eiger 1M fast area detector operating at 10 Hz. A broad diffuse component associated with unit-cell-height islands was consistently observed during film deposition. The specular reflection was split into multiple peaks due to the stepped substrate morphology; these were integrated to obtain the total specular intensity.

The setup was sensitive to in-plane diffuse scattering out to ~3 nm$^{-1}$, corresponding to lateral length scales down to ~2 nm. This enabled detection across the full range of island evolution—from initial nucleation and aggregation through coarsening and coalescence within each monolayer. The diffuse intensity formed nearly circular patterns in reciprocal space, indicating that islands were isotropically distributed across the surface. Additional experimental details are available in a previous publication.[13]

### C. Kinetic Monte Carlo Modeling

Our kinetic Monte Carlo (kMC) approach simulates surface morphology evolution during pulsed laser deposition (PLD) by tracking atomic-scale processes. Each process is assigned an energy barrier E$_i$ and a corresponding rate k$_i$, defined by the Arrhenius relation:

$$k_i = \nu_0 \exp\left(-\frac{E_i}{K_B T}\right) \tag{1}$$

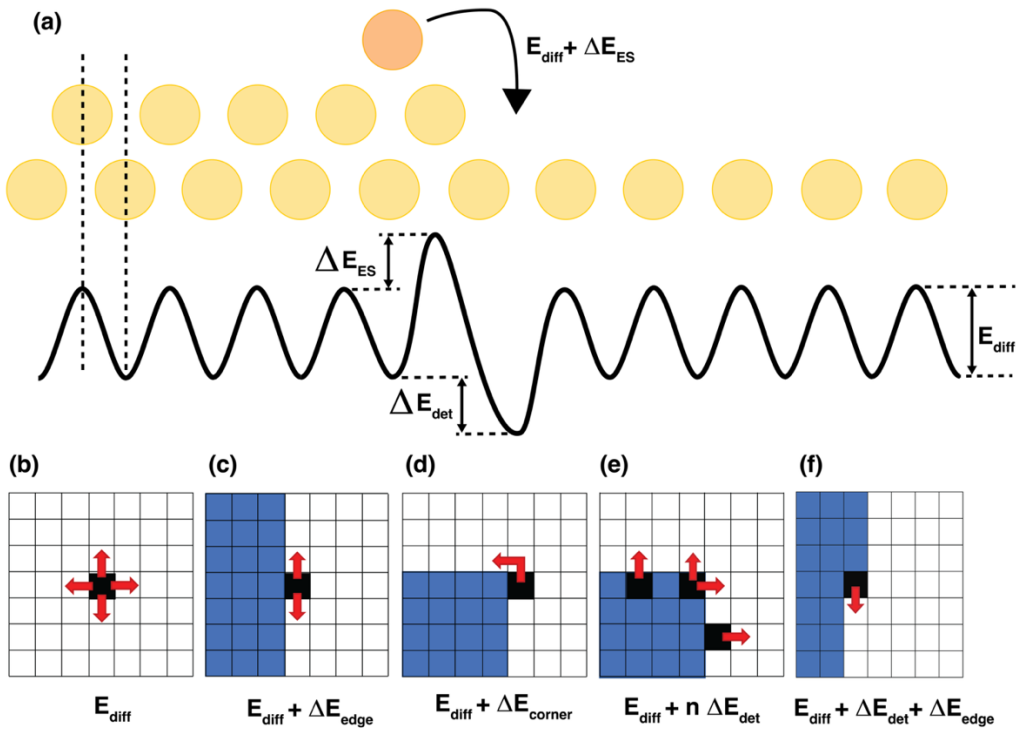

**Figure 2:** Schematic of the energy barriers used in the kMC study. (a) shows is a visualization of the potential energy surface for diffusion of ad-molecules. When the particles reach a step edge, the diffusion barrier is increased by an amount $\Delta E_{ES}$. (b-f) show important processes for intralayer particle diffusion, including surface, edge, and corner diffusion (b-d), as well as step edge (e) and kink detachment (f).

where $\nu_0$ is the attempt frequency, $K_B$ is Boltzmann's constant, T is the absolute temperature. For atomic species, $\nu_0$ is typically $10^{12} - 10^{13}$ Hz, although clusters of several atoms may also act as dominant diffusing species,[6] exhibiting lower attempt frequencies.

All energy barriers are defined relative to the surface diffusion barrier $E_{diff}$ (Figure 2), and simulations are performed at a fixed temperature of 600°C. The surface hopping rate $r_{hop}$ sets the timescale for diffusion and we use it to constrain the choices of $\nu_0$ and $E_{diff}$. For example, with $k_1 \equiv r_{hop} = 407.84$ s$^{-1}$ and a chosen attempt frequency $\nu_0 = 10^{13}$ Hz, $E_{diff}$ corresponds to 1.8 eV (Equation 1). This hopping rate was chosen to match experimental lateral length scales based on preliminary simulation results and is used consistently throughout the study. In the final analysis (Figure 7) we adopted a revised hopping rate of 1251.8 s$^{-1}$ and an attempt frequency of ~$10^7$ Hz, consistent with the experimental results of Ferguson et al.[23] The Erlich Schwoebel barrier ($\Delta E_{ES}$) was set to 0 for all simulations.

Each computational cycle selects a process—either deposition or one of the surface processes illustrated in Figure 2—using a probability weighted by its rate. Time is incremented using the Bortz–Kalos–Lebowitz (BKL) rejection-free algorithm:[24]

$$\Delta t = \frac{-\ln(r)}{\sum_i k_i} \quad , \tag{2}$$

where $r$ is a uniformly distributed random number in (0,1), and the sum is over all active processes. A sorting scheme similar to that of Schulze is used to group identical processes and accelerate updates.[25] PLD is modeled as bursts of deposition followed by dwell periods during which only surface processes occur.

To maintain detailed balance, monomers attached to step edges are allowed to detach upward with an appropriate energy barrier, $E_{det,up} = E_{diff} + n\Delta E_{det} + \Delta E_{ES}$ . In the absence of an Ehrlich-Schwoebel barrier, the detachment rates to upper and lower terraces are equal.

We studied different initial configurations. For circular island simulations, radii ranged from 8 to 33 lattice units (Figures 6 and 7). Recovery after a single deposition pulse was simulated for detachment barrier $\Delta E_{det}$ ranging from 0 and 0.4 eV with at least 800 simulations averaged for each case. Other parameters were fixed at $\Delta E_{ES} = 0$, and $\Delta E_{edge} = \Delta E_{corner} = 10$ eV. For circular islands, upward detachment was disabled to allow better comparison with continuum models. To generate smooth adatom density maps (e.g., Figures 6(f)–(j) with $\Delta E_{det} = 0.3$ eV), 1,000 simulations were used. Additional simulations with varying $\Delta E_{det}$ are described in Supplemental notes 2 and 3.[26]

We also employed a hybrid approach to investigate adatom transport on non-circular island morphologies: a single growth simulation from a smooth starting surface with $\Delta E_{det} = 0.3$ eV was stopped at 0.6 ML coverage and used as an initial configuration. This was followed by 10,000 single-shot simulations — resetting to the initial surface each time — to produce a density map and specular recovery curve [Figures 8(a) and 8(b)]. For Figure 9, growth was simulated from an initially flat surface with 20 monolayers subsequently deposited. To compute the diffuse scattering profiles shown in Figure 9, 100 equivalent simulations were averaged. Scattering calculations are described in the next subsection.

Additional simulation details are provided in Supplemental Notes 1 and 6.[26] A similar kMC implementation was described by Lucas and Moskovkin.[27] The code used for this work is available via a public repository.[28]

## D. Simulated X-ray Scattering Intensities

Both specular and off-specular X-ray scattering patterns were calculated from simulated surfaces represented as height arrays. The specular recovery can be understood in terms of the specular intensity:

$$I_{spec} \sim |F(Q_z)| \left| \frac{1}{1 - e^{iQ_z a}} + \sum_{n=1}^{N} \theta_n(t) e^{-iQ_z na} \right|^2 \tag{3}$$

Note that that at the (00 1/2), which is known as the *specular anti-Bragg* position, we have $Q_z a = \pi$, making the scattering from each layer out of phase with the one beneath it. Typically, only two incomplete layers are present. Then Eq. 3 simplifies to a form involving only the layer coverages $\theta_1$ and $\theta_2$ : $I_{spec} \sim |1 - 2\theta_1 + 2\theta_2|^2$. This expression is used to track the specular intensity during simulated film growth. Individual layer coverages $\theta_n$ evolve with time, so that as material from layer 2 moves to layer 1 the specular intensity increases or decreases for $\theta_1$ coverages greater or less than 0.5, respectively.

The diffuse scattering intensity is computed from the height correlations of the simulated surface. The scalar product of $\vec{Q}$ and $\vec{r}$ is $\vec{Q} \cdot \vec{r} = Q_z h(x, y) + (Q_x x + Q_y y)$, where the second term can be expressed more compactly as $\vec{Q}_{xy} \cdot \vec{r}$. The diffuse scattering intensity is written as:

$$I_{diff}(\vec{Q}_{xy}) \sim \left| \sum_{\vec{r}} e^{i\vec{Q}_{xy} \cdot \vec{r}} e^{iQ_z h(\vec{r})} \right|^2 \tag{4}$$

Since we are mainly interested in the radial profile rather than the two-dimensional scattering pattern, the expression in Eq. 4 is circularly averaged to produce $I_{diff}(Q_r)$. Additional details of the X-ray scattering formalism can be found in Sinha et al., Als-Nielsen and McMorrow, and Evans et al.[29-31]

## E. Continuum modeling of ad-molecule density evolution

To benchmark the kMC simulations, we solved a continuum model assuming non-interacting particles with a concentration represented by the density $\rho(r,t)$. The radially symmetric two-dimensional diffusion equation is:

$$\frac{\partial \rho(r,t)}{\partial t} = D\nabla^2 \rho(r,t) \tag{5}$$

This was solved within a circular domain of radius R, using the initial condition $\rho(r,0) = \rho_0$ and boundary condition $\rho(R) = 0$. As shown in the Supplemental Note 4, depletion occurs first near the edge and then more slowly near the center. Comparison with kMC simulations (for $\Delta E_{det} = 0$ and $\Delta E_{ES} = 0$ ) shows good agreement, with slight differences due to the kMC constraint that prohibits multiple occupancy of lattice sites, leading to a minor delay in particle depletion.

Following Fleet et al.[5], we also modeled diffusion within an annular domain, with a central hole at radius $R_a$ where monomers can drop down, and an outer boundary at $R_b$ representing the average distance between holes. Specular intensity recovery was calculated by integrating the radial concentration profile and applying the kinematical scattering formula. In both cases—circular and annular—the stretching exponent was close to 1.0, indicating that the fundamental mode dominates the relaxation dynamics despite the presence of multiple exponential components. These results are shown as star symbols in Figure 7b. Further details are reported in the Supplemental Note 5.[26]

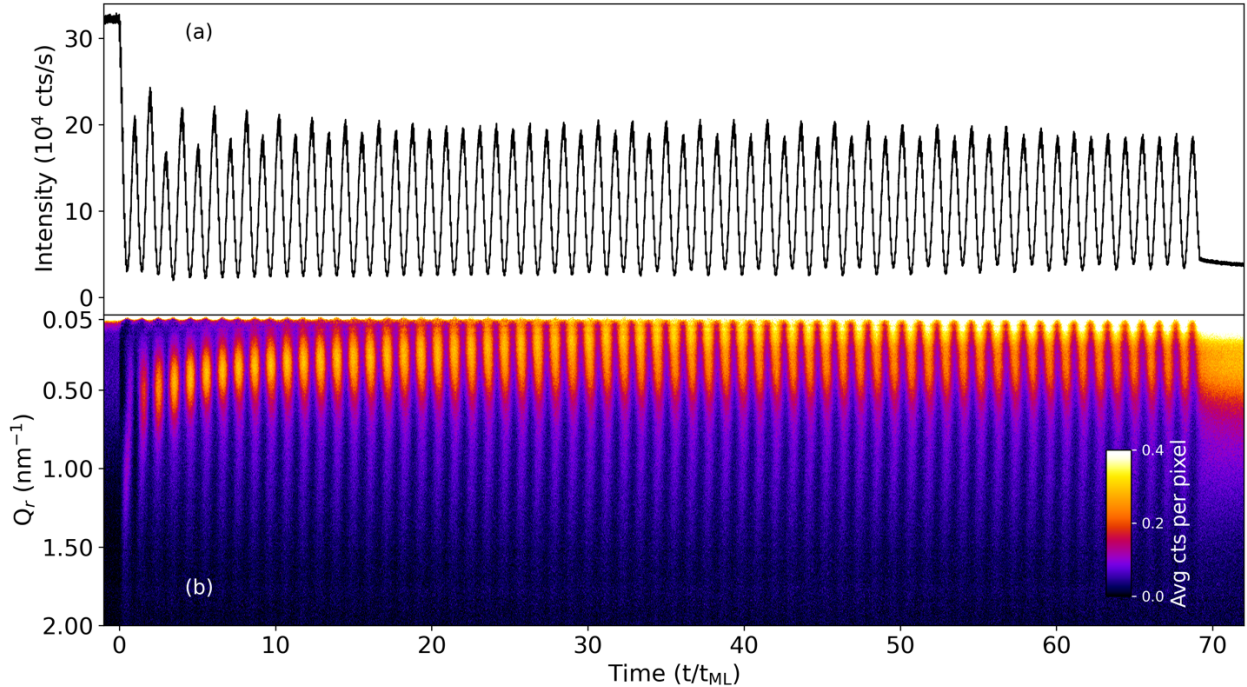

**Figure 3.** (a) Experimental specular anti-Bragg intensity and (b) circularly averaged diffuse intensity over full deposition of 64.5 layers. The horizontal axis has been converted from time to monolayer coverage using 6 sec per laser shot and 18 laser pulses per monolayer.

## III. EXPERIMENTAL RESULTS

Figure 3 shows experimental results capturing the evolution of the film during growth at 600 °C with a 6-second dwell time. In Figure 3(a), the specular anti-Bragg intensity is shown over the deposition of 64.5 monolayers, with approximately 18 laser pulses per layer. The data, acquired at 10 Hz, reveal extended layer-by-layer growth, marked by sharp intensity jumps at each pulse. Figure 3(b) presents the circularly averaged diffuse scattering around the (00 ½) anti-Bragg reflection. In the first deposited layer, the diffuse scattering is peaked near $Q_r$ = 1.0 nm$^{-1}$ corresponds to an in-plane length scale of about 6 nm. Assuming a unit cell spacing of ~0.4 nm, and each pulse deposits 1/18 ML, the mean spacing between monomers is about 1.7 nm. However, the very weak signal at higher $Q_r$ ($\gtrsim 2.0$ nm$^{-1}$) suggests that individual monomers contribute minimally to the diffuse intensity. The dominant peak near 1.0 nm$^{-1}$ arises instead from small islands spaced by approximately 6.3 nm, consistent with the schematic shown in Figure 1. Taking the pulse intensity into account, we estimate they are formed by the aggregation of ~14 unit-cells, corresponding to a linear island size of ~1.5 nm. In subsequent layers, a broader peak near $Q_r$ = 0.5 nm$^{-1}$ emerges and shifts with increasing layer number, while the persistent peak at 1.0 nm$^{-1}$ indicates continued nucleation of small islands during each monolayer cycle.

Figure 4 provides a closer view of both specular and diffuse signals. Figure 4(a) shows the specular intensity during deposition of a single monolayer (Layer 16), with the inset highlighting the region near 0.6 coverage. Red points correspond to frames where deposition pulses occurred. Figure 4(b) displays a single averaged growth oscillation compiled from 60 consecutive layers during the steady-state regime, showing significantly improved statistics. Its inset again zooms in around 0.6 coverage. The specular recovery exhibits a noticeable asymmetry between lower coverages <0.5 ML and higher coverages; we consider this effect in the Discussion section. Figure 4(c) shows the diffuse intensity averaged over Layer 16, fitted with Lorentzian peaks. A broad, high-$Q_r$ shoulder (green dashed line) suggests scattering from small clusters or unresolved features. Figure 4(d) shows the diffuse intensity over one full layer oscillation, averaged over 60 layers. Curves are binned by $Q_r$, corresponding approximately to island size. The blue trace, representing larger islands, increases between pulses as material relaxes downward. In contrast, the green curves represent scattering from smaller clusters, which increase abruptly during deposition and decay between pulses as material transfers to larger islands. This behavior confirms that small islands serve as intermediates in the interlayer transport process.

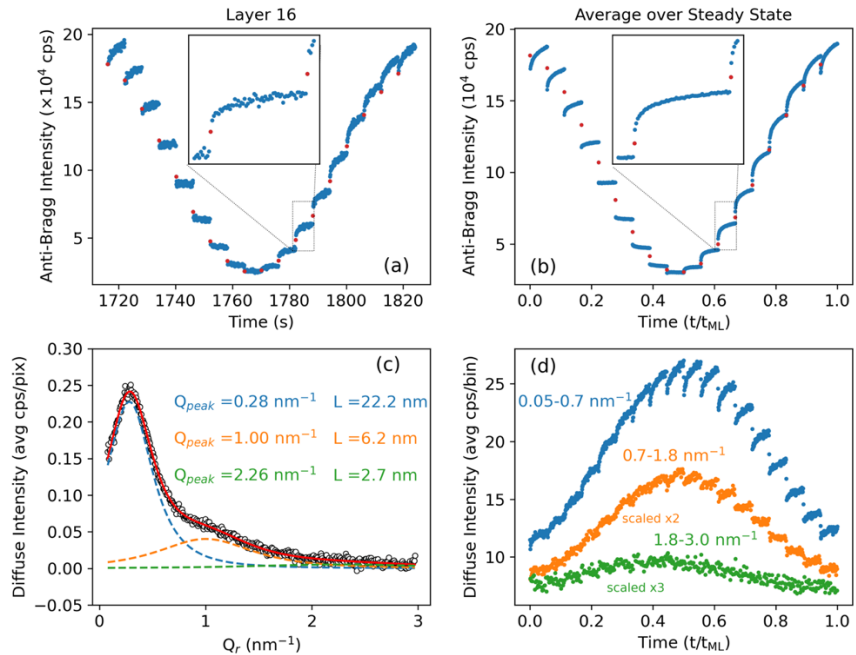

**Figure. 4** Close up views of the experimental specular and diffuse data. (a) One complete monolayer oscillation of the specular intensity. (b) The specular intensity averaged over 60 monolayers. (c) The diffuse intensity averaged over one monolayer. (d) The diffuse scattering averaged over 60 monolayers and further binned over the ranges of $Q_r$ specified in the figure. Details of these curves are discussed in the main text.

Figure 5 shows how the X-ray signal recovers after each laser pulse by fitting the data with a stretched exponential function (as shown in Equation 6), which captures the broad range of relaxation times in the system.

$$I_{spec} \sim I_f + (I_0 - I_f) \exp\left[-\left(\frac{t}{\tau_1}\right)^{\beta_1}\right] \tag{6}$$

Figures 5(a)–(d) provide representative fits at different time intervals, while Figures 5(e) and 5(f) show how the key parameters evolve as the film grows. The relaxation time $\tau_1$ increases with layer coverage, indicating slower relaxation as the layer approaches completion. This effect arises from the increasing influence of longer diffusion distances and longer lasting transient islands as the underlying layer begins to coalesce. The stretching exponent $\beta_1$ remains close to 0.5 in the intermediate regime, reflecting a wide distribution of kinetic pathways.

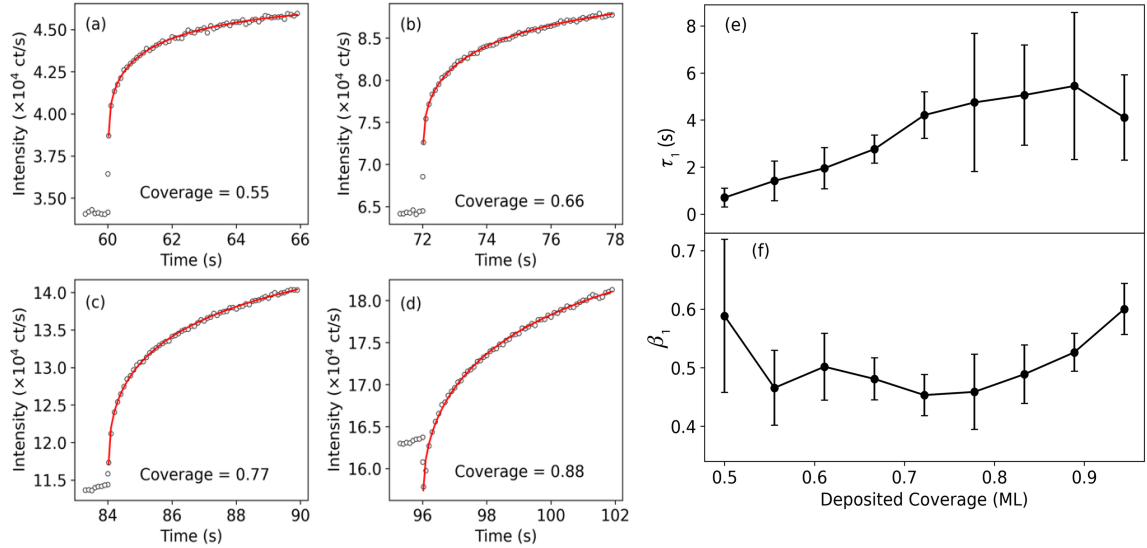

**Figure 5.** Stretched exponential fits to specular anti-Bragg intensity recovery during the second half of each monolayer. Panels (a)–(d) show example fits; panels (e) and (f) plot the extracted parameters. The time constant $\tau_1$ increases with coverage, while the exponent $\beta_1$ rises slightly as islands coalesce and the layer morphology transitions from discrete islands to holes. This trend reflects cancellation effects in the X-ray intensity due to out-of-phase scattering, as described in Equation 3.

## IV. COMPUTATIONAL RESULTS

The kinetic Monte Carlo (kMC) method has been widely used to model thin film growth processes, including simulations of pulsed laser deposition (PLD) and applications to oxide perovskite thin films.[21,32-38] Our implementation is based on a solid-on-solid (SOS) model, in which the surface is represented by a lattice of discrete height values corresponding to monolayers. This approach coarse-grains over individual atomic bonds, treating the mobile species as atoms or small clusters—such as complete $SrTiO_3$ molecular units.

A central strength of kMC is its ability to span a wide range of time scales, enabling the simulation of rapid processes such as surface diffusion and interlayer hopping, alongside slower dynamics like detachment from step edges and island coarsening. Growth conditions—including pulse rate and recovery time—can also be realistically modeled. Each process in the simulation is governed by an activation barrier $E_i$, which determines its probability of occurring. Details of our kMC implementation are described in the Methods section and Supplemental Notes 1 and 6.[26]

Figure 2 summarizes the energy landscape used in the simulations. Figures 2(a) visualizes the potential energy surface for surface diffusion, showing that the barrier increases by an amount $\Delta E_{ES}$ when particles reach a step edge. Figures 2(b)–(f) illustrate key diffusion mechanisms included in the model: surface, edge, and corner diffusion (b–d), as well as step-edge hopping (e) and kink-site detachment (f). In this study, the Ehrlich–Schwoebel barrier for downward hopping is assumed to be negligible ($\Delta E_{ES} \approx 0$), consistent with prior results for $SrTiO_3$. We focus instead on the detachment barrier $\Delta E_{det}$, which is found to control the recovery behavior of both specular and diffuse X-ray scattering during the dwell period after each deposition pulse. The total activation energy for detachment is given by $E_{det} = E_d + n\Delta E_{det}$, where $n$ is the number of lateral bonds that must be broken. Other processes—such as step-edge and corner diffusion, and kink detachment—are disabled in this study (i.e., barriers set to a high value, 10 eV) . We comment on the expected effects of choosing lower values of these parameters in the Discussion section.

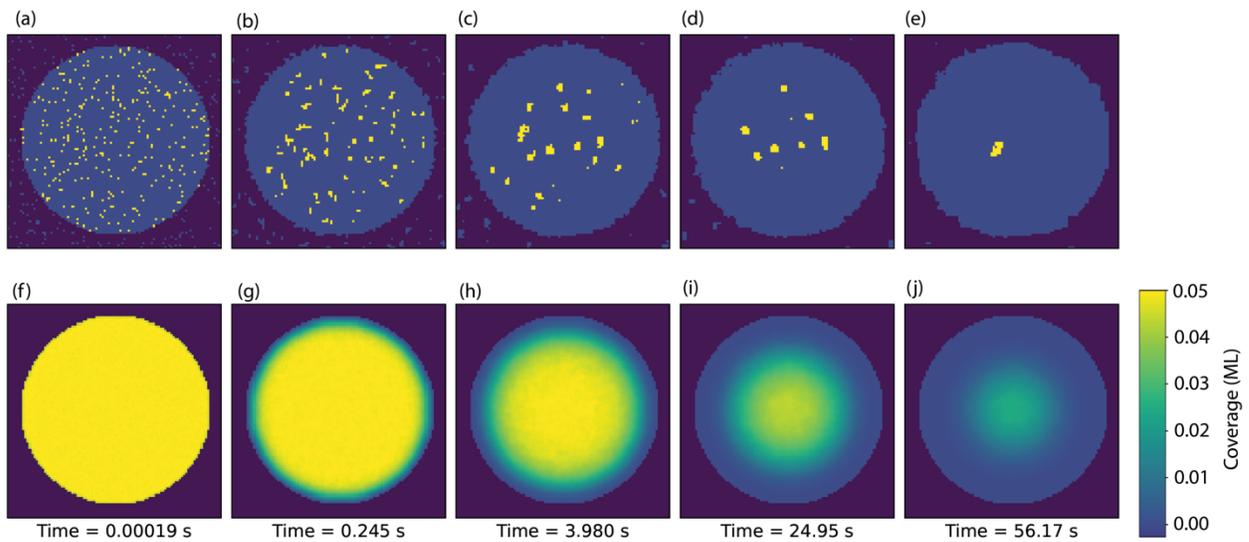

**Figure 6:** Kinetic Monte Carlo study of circular islands with a diameter of 44 lattice units and 0.3 eV detachment barrier. The top row (a-e) is for a single simulation, showing the behavior of transient islands as a function of time after a single deposition pulse with $\Delta\sigma$ = 0.05 ML, and the bottom row (f-j) is an average over 1,000 equivalent simulations. The color scale corresponds to the average height above the base layer, where 0.00 corresponds to the island layer and 0.05 corresponds to 5% coverage on top of the island layer. In (a-f) the transient islands are colored yellow, and the circular lower island is blue on a darker base layer.

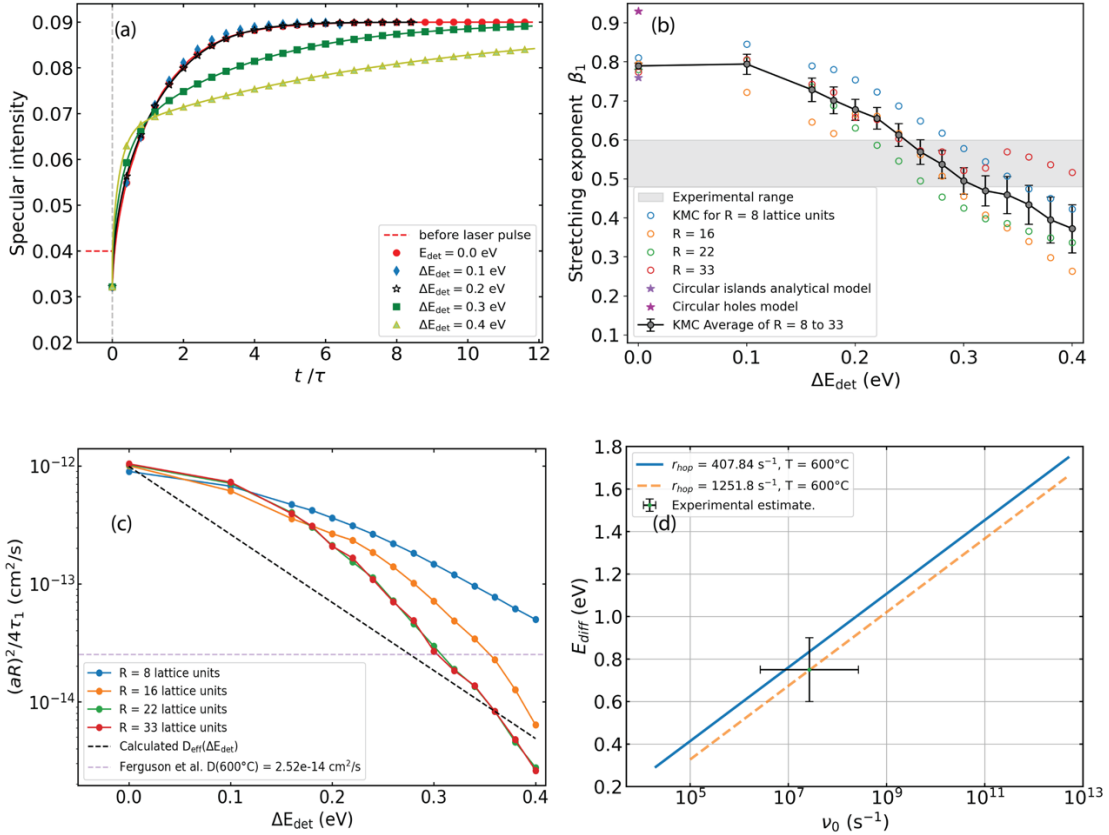

**Figure 7.** Characterization of interlayer transport on circular islands. (a) shows the recovery of the specular intensity as a function of reduced time for islands with a radius of 16 lattice units for several detachment energy barriers. (b) shows results from fitting the recovery curves for several island radii as a function of the detachment energy barrier. (c) shows the estimated effective diffusion parameter $D_{eff}$ derived from the recovery time constant $\tau_1$ and the island radius R. (d) shows a plot of the surface diffusion hopping rate used in this study. The line shows the diffusion energy barrier and the attempt frequency combinations that result in a hopping rate of $r = 407.8$ s$^{-1}$. The data point is an experimentally derived value, as explained in the main text.

To probe the dynamics of transient island formation and dissolution, we first simulated deposition onto prefabricated circular islands. Figure 6 shows the time evolution for islands with a radius of 22 lattice units at 600 °C and a detachment barrier of 0.3 eV. Initially, small clusters form on top of the circular island following a single deposition pulse, but these clusters gradually dissolve due to thermally activated detachment. As time progresses, the number of transient islands decreases, and eventually, only one remains—although it, too, dissolves if the simulation runs long enough. The top row (Figure 6a–e) shows a single simulation, while the bottom row (Figure 6f–j) presents an average over 1,000 equivalent simulations. The widening denuded zone near the island edge highlights the role of step edges as efficient sinks when $\Delta E_{ES}$ is small. In this study, we focus instead on the effect of a non-zero detachment barrier $\Delta E_{det}$, which significantly slows the dissolution of transient islands while allowing freely diffusing ad-particles to reach step edges unimpeded. This energy separation—between the higher detachment barrier and the lower barriers for diffusion and interlayer transport—leads to the stretched exponential behavior observed in both simulations and experiment. Simulations equivalent to Figure 6 but with varying detachment barriers are shown in Supplemental Notes 2 and 3.[26]

Figure 7 quantifies how the detachment barrier $\Delta E_{det}$ influences the recovery of specular intensity. Figure 7(a) shows normalized recovery curves plotted against reduced time, $t/\tau$, where all curves intersect near unity—consistent with stretched exponential kinetics. As $\Delta E_{det}$ increases from 0 to 0.4 eV, the characteristic relaxation time $\tau$ increases by more than a factor of 50, demonstrating the dramatic slowdown caused by even modest detachment barriers. Figure 7(b) shows the corresponding stretching exponent $\beta$, which decreases to values near 0.5 with increasing $\Delta E_{det}$, indicating a broader distribution of relaxation times. The experimental range for $\beta_1$ is matched when $\Delta E_{det} \approx 0.3$ eV, supporting the conclusion that detachment-limited dynamics govern the observed behavior. Figure 7(b) also shows results for analytical models for circular holes or islands (star symbols); see Supplemental Notes 4 and 5.[26]

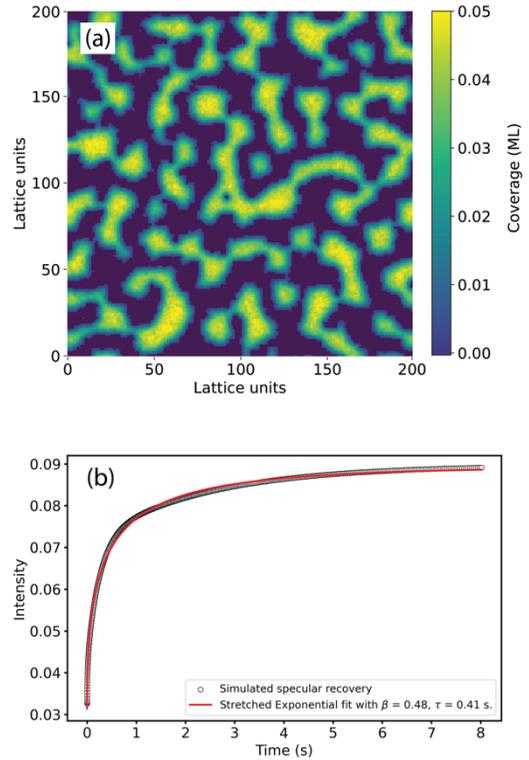

**Figure 8.** Simulated relaxation of a single pulse with $\Delta E_{det} = 0.3$ eV. (a) Ad-molecule density map with a starting coverage of 0.6 ML and 0.05 ML pulse. 0.00 on the color scale represents the island layer. (b) Specular intensity recovery after the pulse.

In Figure 7(c), we convert $\tau$ into an effective diffusion constant, $D_{eff} = L^2/4\tau$, and compare the results with extrapolated experimental data from Ferguson et al. at 600 °C.[23] The simulation data align well with this prior work for larger island sizes when $\Delta E_{det} \approx 0.3$ eV. The dashed line in Figure 7(c) represents a crude model based on Arrhenius behavior, given by Equation 7:

$$D_{eff}(\Delta E_{det}) = D_0 \exp\left(-\frac{E_{diff} + \Delta E_{det}}{K_B T}\right) = D_{eff}(0)e^{-\frac{\Delta E_{det}}{K_B T}} \tag{7}$$

which assumes a constant base activation energy $E_d$ for diffusion and isolates the effect of increasing detachment energy. Note that $D_{eff}(0) = D_0 \exp(-E_{diff}/K_B T)$ is simply the monomer diffusion rate on the free surface neglecting the effect of attachment/detachment kinetics. This model captures the overall trend and highlights the exponential sensitivity of recovery kinetics to even small increases in $\Delta E_{det}$. Figure 7(d) further illustrates the relationship between activation energy, attempt frequency, and hopping rate. The experimentally derived point corresponds to a surface diffusion barrier of $E_{diff}$ = 0.75 eV (see the Discussion section for details of the calculation). A reference curve (yellow dashed line) that covers a wide range of attempt frequencies for $r_{hop}$ = 1251 s⁻¹ is shown. The other reference curve corresponds to a constant hopping rate of 407.8 s⁻¹, which we have used in our kMC simulations.

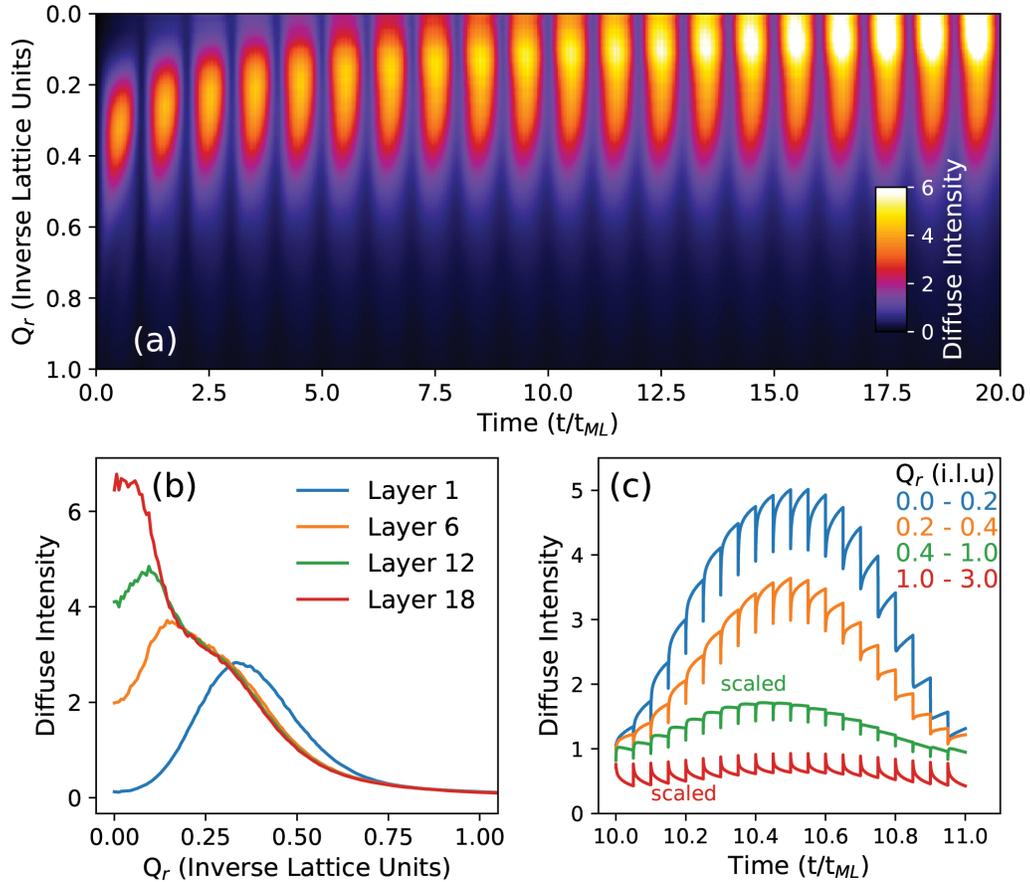

**Figure 9.** Simulated X-ray diffuse scattering for PLD growth. (a) Diffuse scattering vs. time of deposition over 20 monolayers. (b) Time slices integrated over a complete monolayer of growth for layers 1, 6, 12 and 18. (c) Time slices at different $Qr$ highlighting the different relaxation behavior for high and low $Qr$. High $Qr$ corresponds to smaller features.

To test how these results extend to more realistic island morphologies, we performed simulations on a surface with 0.6 ML initial coverage and applied a 0.05 ML pulse. Figure 8(a) shows the ad-molecule density at 0.171 s after the pulse. The initial morphology was generated by standard PLD growth over a 1000×1000 lattice with 6 sec. relaxation after each pulse, and a 200×200 subregion was analyzed. Averaging over 10,000 simulations shows that the denuded zone follows island contours, just as it did for the case of circular islands.

Figure 8(b) shows the specular intensity recovery, which exhibits a stretched exponential shape with β = 0.48, matching experimental observations. The time constant is significantly longer than would be expected in the absence of a detachment barrier, confirming the dominant role of transient island dissolution. These results demonstrate that detachment-limited relaxation is largely independent of island morphology, except in extreme cases where clusters cannot form or coalesce.

Finally, Figure 9 shows the simulated diffuse X-ray scattering from kMC simulations of 10-layer PLD growth using 0.05 ML pulses and 6 s dwell time. In Figure 9(a), the diffuse scattering is circularly averaged in $Q_r$ to obtain the radial dependence. The results closely match the experimental patterns shown in Figure 2. Notably, the initial layers show a dominant short length scale component (peaked at high $Q_r$), which gradually evolves into a longer length scale feature as layer coalescence progresses. Figure 9(b) highlights this evolution by comparing the diffuse scattering integrated over full monolayers for layers 1, 6, 12, and 18. A shoulder at high $Q_r$ persists across layers, indicating continued nucleation of small features atop the growing film. Note that if we convert the $Q_r$ scale in Figure 4(c) using the conversion 1 nm$^{-1}$ = 0.39 inverse lattice units (ilu), we see that the scales are similar, and the high $Q_r$ shoulder is at a similar position. Figure 9(c) illustrates the recovery behavior after each pulse of a full monolayer deposition (layer 11). The statistics are much better than the experimental data in Figure 4(d), particularly at high $Q_r$. Notably, we see that the high $Q_r$ component decays downward immediately after each deposition pulse, consistent with the dissolution of transient islands. In contrast, the lower $Q_r$ intensities recover upwards during each recovery interval as material is added to larger islands. This effect is described in more detail in the Discussion section. The overall diffuse intensity gradually diminishes at the end of each monolayer growth cycle as the larger islands coalesce, and the surface returns to a relatively smooth state. These trends capture the influence of the underlying morphology on the development of each layer. Additional results, including showing the effect of varying detachment barriers are given in Supplemental Notes 6 and 7.[26]

## V. DISCUSSION

Our kMC simulations reveal that the effective surface diffusion coefficient extracted from X-ray scattering, $D_{eff}$, is systematically lower than the true monomer diffusion coefficient, $D$. This discrepancy arises because transient island formation temporarily traps monomers, delaying interlayer relaxation. Attachment and detachment kinetics introduce an intermediate timescale in specular recovery, reducing the apparent diffusivity extracted from time-resolved data. In this section, we place our findings in the context of prior SrTiO$_3$ PLD studies and demonstrate quantitative consistency by comparing kMC stretched exponential with bi-exponential analyses.

## A. Bi-exponential analysis

Although our stretched-exponential fits require one fewer free parameter and generally yield superior statistics for $\Delta E_{det}$ < 0.5 eV, a bi-exponential fit provides a transparent method for extracting two relaxation times, $\tau_{1a}$ and $\tau_{1b}$, directly from the experimental recovery curves (see Supplemental Note 8).[26] As in Figure 1, $\tau_{1a}$ reflects the (mostly) unimpeded monomer diffusion timescale, while $\tau_{1b}$ is strongly affected by detachment kinetics. Following Gabriel et al.,[21] we model these times with simple Arrhenius expressions:

$$\tau_{1a}^{-1} = \nu_0 \exp\left(-\frac{E_{diff}}{K_B T}\right);$$ (8a)

$$\tau_{1b}^{-1} = \nu_0 \exp\left(-\frac{E_{diff} + \Delta E_{det}}{K_B T}\right).$$ (8b)

Using a conventional attempt frequency $\nu_0 = 10^{11}$ s$^{-1}$, Eq. 8a yields $E_d \approx 2.03$ eV, nearly identical to the 2.05 eV reported by Gabriel et al.[21] However, this approach neglects the diffusion length, $L_d$, that deposited particles must traverse before reaching a step edge. Since $D_0 = \frac{1}{4} a^2 \nu_0$ in two dimensions,[39] and $D \approx L_d^2/4\tau_{1a}$, the left side of Eq. 8a should be rescaled by $(L_d{}^2/a^2)$, yielding an estimate of the monomer hopping rate:

$$r_{hop} = \nu_0 \exp\left(-\frac{E_{diff}}{K_B T}\right) \approx \frac{L_d{}^2}{a^2 \tau_{1a}}.$$ (9)

Estimating $L_d$ from the peak position for layer 2 in Figure 3(b) ($Q_{peak} \approx 0.50$ nm$^{-1}$) gives $L_d \approx \pi/Q_{peak} \approx 6.3$ nm or $\sim$16 lattice constants. Independent temperature-dependent measurements by Ferguson et al.[23] yield $D_0 \approx 10^{-8}$ cm$^2$/s, which corresponds to $\nu_0 \approx 2.67 \times 10^7$ s$^{-1}$ for $a$ = 3.926 Å. Using our experimental $\tau_{1a} \approx 0.20$ s, we obtain $r_{hop} \approx 1.25 \times 10^3$ s$^{-1}$, $E_{diff} \approx 0.75 \pm 0.15$ eV.

This value of $E_{diff}$, directly extracted from experiment, agrees well with our kMC-derived estimate based on stretched exponential fits (see below). It is, however, significantly lower than the uncorrected value of 2.03 eV, highlighting the critical role of transient trapping and detachment. Even with $\nu_0 = 10^{11}$ s$^{-1}$, the diffusion barrier estimated from Eq. 8a is 1.37 eV, still well below the uncorrected estimate.

## B. Consistency with kMC

Our kMC simulations were run with a fixed hopping rate $r_{hop}$ = 407.8 s$^{-1}$, equivalent to $E_d$ = 0.83 eV for $\nu_0 = 2.67 \times 10^7$ s$^{-1}$ at 600 °C. Although we did not fine-tune $r_{hop}$ to match the experiments perfectly, the model demonstrates internal consistency. For example, Figure 7(c) demonstrates that when $\Delta E_{det}$ = 0, data for different circular island sizes collapse onto a single $L^2/4\tau$ point—confirming that the kMC simulations correctly capture the effective diffusion length.

As an additional benchmark, a detailed comparison of circular island results for the kMC model vs a continuum model is provided in Supplemental Note 4.[26]

The good agreement between simulation results (Fig. 9 and Supplemental Fig. 11)[26] and experiment (Fig. 4) is particularly notable given the coarse-grained nature of the kMC model.

The simulations assume that deposition results in complete unit cells, effectively neglecting local variations such as half-unit-cell growth (e.g., $TiO_2$ or $SrO$ termination) or amorphous deposits. This approximation is justified if crystallization occurs on a timescale faster than surface diffusion to step edges and interlayer transport. The agreement with experiment suggests that more complex crystallization pathways either play a negligible role or occur at length and time scales not resolved in the measurements.

## C. Extracting $\Delta E_{det}$ from experiment

A more direct determination of the detachment barrier follows from the ratio of Eqs. 8a and 8b:

$$\frac{\tau_{1b}}{\tau_{1a}} = exp\left(\frac{\Delta E_{det}}{K_B T}\right) \tag{10}$$

Using our measured ratio of approximately 14 at 600 °C yields $\Delta E_{det} \approx$ 0.20 eV, consistent with the results of Gabriel et al. , and slightly lower than our kMC estimate of 0.30 ± 0.05 eV. We consider the kMC estimate to be more accurate because it quantitatively accounts for both the recovery line shape [Figure 7a,b] and the diffusion coefficient correction [Figure 7c]. Figure 7c clearly predicts that the effect of the detachment barrier is to suppress the effective diffusion coefficient by a factor of ~50 rather than 14. The measurement of $\tau_{1a}$ from biexponential fitting may be an overestimate because it may not be purely a measure of the barrier-free monomer time constant.

We note that a step edge barrier $\Delta E_{ES} > 0$ alone (i.e. along with $\Delta E_{det}$=0) will produce a different effect since for $\Delta E_{ES} \neq 0$ interlayer transport will be delayed even for particles that did not attach to transient islands. Thus, it would not lead to stretched exponential or bi-exponential recovery of the specular intensity. On the other hand, the addition of a very small $\Delta E_{ES}$ in combination with a larger $\Delta E_{det}$ would be expected to delay the fast component of the recovery only, leaving the slow part relatively unchanged.

## D. Comparison to activation energies from temperature studies

Ferguson et al. varied the PLD growth temperature to extract both $D_0$ and an effective activation energy, reporting $E_a$ = 0.97 ± 0.07 eV and $D_0 \approx 10^{-8}$ cm$^2$/s$^{-1}$. A simple exponential was used to extract the recovery time constants, which we assume is mostly characteristic of the delayed $\tau_{1b}$. Applying Eq. 7, we find $E_a = E_{diff} + \Delta E_{det}$. Subtracting our detachment barrier, $E_{diff}$ = 0.67 ± 0.09 eV, in excellent agreement with our combined experimental and simulation-based estimates.

Collectively, these results demonstrate that detachment-limited interlayer transport dominates the long-time relaxation dynamics in $SrTiO_3$ PLD. Both stretched- and bi-exponential models yield a consistent physical picture: while monomer diffusion sets the fundamental timescale, transient trapping at step edges governs the observed relaxation behavior.

### E. Coverage dependence of specular and diffuse recovery

The specular intensity shown in Fig. 4(a, b) exhibits a distinct asymmetry between the first and second halves of the monolayer. In the early stage of the layer ($\theta_1 < 0.5$), corresponding to island nucleation and early aggregation, the post-pulse intensity recovery is nearly flat or even slightly decreasing. An exception occurs at very low coverage ($\theta_1 \leq 0.1$ ML), where the deposited material primarily fills residual holes left by incomplete coalescence of the previous layer.

In contrast, when islands begin to coalesce for $\theta_1 > 0.5$ the recovery becomes strongly upward, driven primarily by interlayer mass transport. For example, the specular intensity at ~0.8 ML (just prior to coalescence) shows pronounced relaxation, unlike the behavior observed at ~0.2 ML. This difference is attributed to the evolution of island size and stability: at lower coverage, small, stable islands dominate, whereas larger, coalescing islands appear at higher coverage.

We interpret the asymmetry as arising from two main effects: (i) Reduced particle capture by small islands since they cover a small percentage of the surface area, leading to weaker observed relaxation; and (ii) Inability of small islands to support transient second-layer islands, since they tend to "shed" deposited material before aggregates can stabilize. These mechanisms are corroborated by the simulated specular intensity results (Supplemental Fig. 11) .[26]

The total diffuse intensity behaves complementarily to the specular signal. As shown in Fig. 3(a, b), the two signals oscillate out of phase—diffuse intensity peaks when the specular reaches a minimum near half-integer monolayer coverages. However, the $Q_r$-resolved diffuse recovery reveals more complex behavior due to its sensitivity to in-plane mass redistribution associated with island aggregation and coarsening.

Fig. 4(d) illustrates this complexity: the $Q_r$ = 0.05–0.7 nm$^{-1}$ (blue) curve shows an upward relaxation after each deposition pulse throughout the full monolayer cycle, indicating net growth at longer length scales. At higher $Q_r$, however, this trend reverses. For example, diffuse features in the range $Q_r$ = 1.8–3.0 nm$^{-1}$ (green curve corresponding to real-space features ~2.1–3.5 nm) exhibit a downward trend, suggesting that these small-scale features shrink or merge into larger structures during recovery.

These coarsening dynamics, both the growth of larger features and the decay of smaller ones, are also evident in the KMC simulations [Fig. 9(c)], where the signal-to-noise ratio is higher and the trends appear more clearly. The diffuse intensities are resolved into Qr bins that roughly correspond to the experimental results, highlighting the similarities.

## F. Multilayer coarsening behavior

In-plane coarsening during multilayer PLD growth involves two distinct populations of surface features that evolve on different timescales. The first consists of small, transient islands that repeatedly nucleate and dissolve during the recovery time following each deposition pulse. The second comprises a population of larger, stable islands that persist and coarsen gradually across many monolayers.

Evidence for the transient island population comes from the persistent appearance of high-$Q_r$ diffuse scattering immediately after each laser pulse (Figures 3, 4). These proto-islands, estimated to be as small as 1.5 nm (Figure 4d), generate observable diffuse scattering extending to $Q_r \approx 3.0$ nm$^{-1}$. Their rapid dissolution is driven by non-equilibrium, detachment-limited kinetics: monomers detach and either hop down to lower layers (interlayer transport) or diffuse to larger islands (intralayer transport). Because monomers that reach step edges are quickly incorporated, they are effectively removed from the local population, resulting in fast, irreversible relaxation. This explains why transient islands evolve on sub-second to few-second timescales within each dwell period.

In contrast, a second diffuse scattering component emerges at lower $Q_r$, reflecting the slower coarsening of a more stable island population. This component begins to form around 2.5 monolayers and becomes increasingly well-defined with continued growth. By layer 16, the dominant diffuse peak stabilizes near $Q_r \approx 0.28$ nm$^{-1}$, corresponding to island diameters of ~11 nm or 28 lattice constants (Figure 4c). This population continues to coarsen as additional layers are deposited, with the peak position gradually shifting to lower $Q_r$ values. This regime operates closer to equilibrium, involving mass exchange between islands of comparable size over many deposition cycles.

Further support for the existence of a slow coarsening regime comes from the work of Eres et al., who observed that under static conditions—i.e., when deposition is halted—coarsening effectively stops at ~2.5 monolayers.[8] No measurable lateral redistribution occurred during annealing at 640°C, despite the presence of islands with diameters of ~10 nm. This observation aligns well with our own findings at layer 16, where stable islands reach similar sizes (~11 nm) and the diffuse scattering peak begins to stabilize near $Q_r \approx$ 0.28 nm$^{-1}$ (Figure 4c).

The lack of further coarsening in the Eres experiment is consistent with expectations from the Gibbs–Thomson effect, which governs Ostwald ripening: once small islands dissolve and a relatively narrow size distribution remains, the chemical potential difference that drives mass transfer between islands becomes minimal. As a result, the coarsening process slows dramatically or becomes effectively arrested. This distinction underscores the contrasting dynamics of transient islands, which evolve rapidly because they are far from equilibrium, versus stable islands, which evolve slowly under near-equilibrium conditions.

Our kMC simulations capture both regimes. As shown in Figure 9, the main diffuse peak shifts steadily toward lower $Q_r$, reflecting the slow growth of stable islands, while a broad high- $Q_r$ feature reappears after each pulse, confirming the repeated formation and dissolution of transient islands during recovery. A full treatment of this multilayer coarsening behavior is beyond the scope of the present study, but the observed trends underscore the rich interplay between non-equilibrium processes that shape thin film morphology during PLD.

## VI. CONCLUSIONS

By correlating experimental measurements with kinetic Monte Carlo simulations, this work provides new insight into the mechanisms governing interlayer and intralayer transport during pulsed laser deposition of $SrTiO_3$. A central finding is the critical role of step-edge detachment in shaping growth dynamics, leading to strongly nonequilibrium evolution of small islands during the recovery period following each laser pulse. The monomer diffusion length—determined by island size and density—is equally important, as it helps set the timescale of interlayer relaxation.

Thermally driven interlayer transport occurs in two distinct stages: an initial fast relaxation, as monomers landing near step edges incorporate directly, followed by a slower process dominated by detachment from transient islands and diffusion across the terrace. This nonequilibrium behavior is governed by the interplay of both diffusion and detachment kinetics. We show that the early-time recovery is diffusion-limited—determined by the monomer hopping rate and largely independent of the detachment barrier—while the later-time behavior is governed by the rate of monomer reemission from step edges and strongly influenced by the detachment energy barrier.

In $SrTiO_3$, where Erlich-Schwoebel step-edge barriers are known to be small, transient islands tend to nucleate near the centers of underlying stable islands rather than uniformly across the surface. This nonuniform nucleation pattern likely influences long-term surface morphology, including lateral coarsening over multiple monolayers. These findings underscore the importance of incorporating transient island dynamics, detachment kinetics, and diffusion length into models of PLD growth to accurately describe relaxation and morphological evolution.

### DATA AVAILABILITY

The data that support the findings of this article are openly available.[40]

### ACKNOWLEDGEMENTS

The authors acknowledge the contributions of K. Evans-Lutterodt. This material is based on work that was supported by the National Science Foundation under Grant No. DMR-